\documentclass[a4paper,11pt]{article}

\usepackage{bm}
\usepackage{latexsym}
\usepackage{amssymb}
\usepackage{amsmath}
\usepackage{mathtools}
\usepackage{hyperref}
\hypersetup{colorlinks=true, allcolors=blue}
\usepackage{graphicx}

\usepackage{mathbbol}
\usepackage{braket}
\usepackage{empheq}
\usepackage{cite}
\usepackage[title]{appendix}
\usepackage{here}
\usepackage{pifont}

\makeatletter
\renewcommand\subparagraph{
    \@startsection {subparagraph}{5}{\z@ }{3.25ex \@plus 1ex
    \@minus .2ex}{-1em}{\normalfont \normalsize \bfseries }}
\makeatother

\numberwithin{equation}{section}

\begin{document}
\pagestyle{empty}

\vspace{-4cm}
\begin{center}
    \hfill KEK-TH-2850, YITP-26-80 \\
\end{center}

\vspace{2cm}

\begin{center}

{\bf\LARGE
Coherent collective response in many-qubit systems for dark matter detection
}

\vspace*{1.5cm}
{\large 
Ryuichiro Kitano$^1$ and Ryoto Takai$^{1,2,3}$
} \\
\vspace*{0.5cm}

{\it 
$^1$Yukawa Institute for Theoretical Physics, Kyoto University,
Kyoto 606-8502, Japan \\
$^2$KEK Theory Center, Tsukuba 305-0801, Japan\\
$^3$The Graduate University for Advanced Studies, SOKENDAI,
Tsukuba 305-0801, Japan\\
}

\end{center}

\vspace*{1.0cm}

\begin{abstract}
{\normalsize \noindent 
We propose an array of Ramsey-type interferometers using $N$
superposition states,  $(\ket{0} + \ket{1})^{\otimes N}$, as a sensor
to detect wave-like dark matter.
After exposure to the dark matter wave, which induces coherent
qubit transitions, the signal is the imbalance between the numbers
of 0 and 1 outcomes.
The signal-to-noise ratio in this scheme is proportional to $N \alpha^2$,
where $\alpha$ is the coupling of dark matter to the qubits, and thus the
sensitivity to the coupling scales as $\delta \alpha \sim 1 / \sqrt{N}$.
For comparison, in the detection scheme based on the Rabi-type
transition, $\ket{0} \to \ket{1}$, this scaling is achieved only
when $N$ highly entangled qubits are used.
Since the Ramsey-type measurement does not require entangled states,
one can consider much larger $N$ by simply placing a large number of
qubits within the de Broglie wavelength of the dark matter.
We demonstrate that, using trapped-ion qubits in linear Paul traps as
the sensor, the projected sensitivity to the coupling matches or surpasses
existing laboratory, astrophysical, and cosmological bounds for $N
\gtrsim 10^6$--$10^8$.
We also evaluate its sensitivity to high-frequency gravitational waves.
Our general framework should, in principle, be useful for other quantum
sensing platforms.
}
\end{abstract} 


\newpage
\baselineskip=18pt
\setcounter{page}{2}
\pagestyle{plain}

\setcounter{footnote}{0}

\tableofcontents
\noindent\hrulefill


\section{Introduction}

The nature of dark matter remains unknown despite overwhelming
gravitational evidence for its existence~\cite{Zwicky:1933gu,
Zwicky:1937zza,Clowe:2003tk,Markevitch:2003at,Planck:2018vyg}.
Among the many proposed candidates, ultralight bosons, including
axion-like particles and dark photons, have attracted considerable
attention~\cite{Arias:2012az}.
When the dark matter mass is sufficiently small, the occupation number
within a de Broglie volume is much larger than unity, and the dark matter
field is well described as a coherent oscillating background within its
coherence time, with an oscillation frequency determined by its mass.

Recent advances in quantum technologies have opened new frontiers in
precision measurement and sensing~\cite{Degen:2016pxo}, with platforms
such as superconducting circuits~\cite{Dixit:2020ymh,Chen:2022quj,
Chen:2024aya}, Rydberg atoms~\cite{Engelhardt:2023qjf,Chigusa:2025rqs},
cavities~\cite{Lamoreaux:2013koa,HAYSTAC:2020kwv,Cai:2025fpe,Freiman:2025tse},
and ion traps~\cite{Gilmore:2021qqo,Ito:2023zhp,Ito:2025mgm}
being actively explored.
When dark matter interacts with such systems, the quantum state is
slightly modified, and this change can be detected as a signal of dark
matter.
Since each platform is sensitive to a different mass range,
complementary operation of different platforms is desirable.

The simplest scheme based on platforms of quantum computers which employ
qubits, a quantum analog of classical bits, is as
follows~\cite{Chen:2022quj,Ito:2023zhp}.
One first prepares the state $\Ket{0}$ as an initial state, and
measures the qubit state after a certain waiting time.
The signal is a non-zero probability of detecting 1, arising from
resonant excitation of the qubits when the qubit transition frequency
is tuned to the dark matter mass.
The signal is proportional to $\alpha^2$ for a single qubit, with
$\alpha$ denoting the coupling of dark matter to the qubit.
Normally, using $N$ qubits enhances the signal by a factor of $N$
compared with the single-qubit case, so that the signal-to-noise ratio
of this Rabi-type scheme is given by ${\rm SNR} \sim \sqrt{N} \alpha^2$,
assuming the noise rate increases linearly in $N$.
The sensitivity to $\alpha$ is determined by ${\rm SNR} = 1$ and
obtained as $\delta \alpha \sim N^{-1/4}$.

Quantum technologies, such as entanglement, squeezing, and quantum error
correction, provide detection schemes that achieve sensitivities beyond
this $N$-scaling or that reduce background events~\cite{Chen:2023swh,
Ito:2023zhp,Takai:2025cyy,Chen:2025tgj,Fukuda:2025afi,Nakano:2026gxq,
Tan:2026poz}.
For example, maximally entangled states of qubits enable the
enhancement of the signal rate by a factor of
$N^2$~\cite{Chen:2023swh,Ito:2023zhp}.
The signal-to-noise ratio scales as ${\rm SNR} \sim N^{3/2} \alpha^2$
when the noise rate is proportional to $N$ or ${\rm SNR} \sim N \alpha^2$
when the noise rate is proportional to $N^2$.
Although sophisticated operations are expected to become feasible with
future technological advances, it remains challenging to implement them
with current technology for large $N$~\cite{Bruzewicz:2019jik}.

On the other hand, it is well known that a superposition state of a qubit
is sensitive to the relative phase rotation between $\Ket{0}$ and
$\Ket{1}$.
The state $\Ket{0} + \Ket{1}$ is initially prepared using the Hadamard
gate, $\Ket{0} \to \Ket{0} + \Ket{1}$ and $\Ket{1} \to \Ket{0} -
\Ket{1}$.
After exposing the qubit to dark matter, the probability of detecting 1 is
given by $1/2 + {\cal O} (\alpha)$, which exhibits better scaling for
small $\alpha$ than the Rabi-type measurement.
However, since the ${\cal O} (\alpha)$ term depends on the unknown phase
of dark matter and vanishes when taking its average, this Ramsey-type
measurement with a single qubit does not help in detecting dark
matter~\cite{Fukuda:2025afi}.

The situation is different for large $N$.
Fortunately, dark matter acts on an array of qubits coherently within
the de Broglie wavelength, enabling a comparison of the probabilities of
detecting 0 and 1 in a single measurement.
Dark matter leaves its trace as the imbalance between 0 and 1
measurements, i.e., $|N_0 - N_1|$, whose excess over the no-signal
expectation is proportional to $N^{3/2} \alpha^2$.
The statistical uncertainty is of order $\sqrt{N}$.
Errors from state preparation, state readout, and heating also scale as
$\sqrt{N}$ if they are independent across the qubits. Mean offsets in
the population imbalance can be subtracted using calibration
measurements, while the residual calibration uncertainty must remain
below the projection noise. Under these assumptions,
the signal-to-noise ratio is ${\rm SNR} \sim N \alpha^2$.

Note that this method does not require entanglement among the qubits.
Thus, the qubits need not all be contained in the same quantum device,
although synchronization of the single-qubit operations as well as the precise
tuning of relevant transition frequencies are necessary.
In such a case, the protocol should be much easier to scale up than
schemes using entangled qubits.

In this work, we consider the dark matter detection scheme
based on this Ramsey-type measurement. We
employ linear Paul traps as an example platform and
estimate the projected sensitivity to the couplings of axion and dark
photon dark matter in the neV mass range to the photon.
These sensitivities can match or surpass existing laboratory,
astrophysical, and cosmological bounds for a large enough number of
qubits, such as $N \sim 10^6$--$10^8$.
In addition, linear Paul traps can search for gravitational waves in the
MHz frequency range for which a similar protocol applies, and the
sensitivity is estimated in Appendix~\ref{sec:gw}.


\section{Experimental setup}

The linear Paul trap has emerged as one of the leading platforms for
quantum computation~\cite{Bruzewicz:2019jik}, and its application to
quantum sensing for wave-like dark matter~\cite{Ito:2023zhp} and
high-frequency gravitational waves~\cite{Takai:2025cyy} has been discussed.
First, we briefly review the structure of linear Paul traps and
the quantum operations used for computation.
In the subsequent subsection, a new detection protocol is proposed and
compared with other protocols in terms of sensitivity.


\subsection{Linear Paul traps}

In the linear Paul trap, singly charged ions, typically beryllium,
calcium, barium, or ytterbium, are held in an ultrahigh-vacuum chamber
through a combination of static and oscillating electric fields, forming
a one-dimensional chain along the trap axis ($z$ axis).
The internal structure of each ion provides an effective two-level system
through optical, fine-structure, hyperfine, or Zeeman transitions.
These systems serve as spin qubits, with $\Ket{\downarrow}$ and
$\Ket{\uparrow}$ labeling the ground and excited states, respectively.

A key operational principle of Paul traps is the coupling between the
ions' internal states and their quantized collective
motion~\cite{Cirac:1995zz}.
The inter-ion spacing, determined by the equilibrium between Coulomb
repulsion and the confining potential, is typically of order
$1\text{--}10$~{\textmu}m.
Following laser cooling, the motional degrees of freedom are well
described by a set of quantum harmonic modes;
in this work, we restrict our attention to oscillations along
the $z$ direction.
For the single-ion case, the coupled spin-motional system is governed by
the free Hamiltonian
\begin{equation}
    \hat{H}_0 = \frac{\omega_0}{2} \hat{\sigma}_z + \omega_z
    \hat{a}^\dagger \hat{a} ,
\end{equation}
where $\omega_z$ ($\sim$ MHz) and $\omega_0$ ($\sim$ GHz) denote
the motional and spin-transition frequencies, respectively.
The two lowest Fock states $\Ket{n = 0}$ and $\Ket{n = 1}$ define a
vibrational qubit, and $\hat{a}^{(\dagger)}$ are its ladder operators.
The Pauli matrix $\hat{\sigma}_z$ acts on the spin qubit.

Precise control over the ion's quantum state is achieved by driving laser
fields at frequencies resonant with distinct sideband transitions.
In the interaction picture and within the Lamb--Dicke approximation
($\eta_{\rm LD} \sim 0.01\text{--}0.1$), the ion-laser
interaction~\cite{walls2008quantum}
\begin{equation}
    \hat{H}_{\rm laser} = \frac{\Omega}{2} \hat{\sigma}_+ e^{-i (\omega -
    \omega_0) t + i \phi} + \frac{i}{2} \eta_{\rm LD} \Omega \,
    \hat{a}^\dagger \left( \hat{\sigma}_+ e^{-i (\omega - \omega_0 -
    \omega_z) t + i \phi} - \hat{\sigma}_- e^{i (\omega - \omega_0 +
    \omega_z) t - i \phi} \right) + {\rm h.c.}
\end{equation}
enables three selective transitions: the red sideband ($\Ket{\downarrow,
1} \leftrightarrow \Ket{\uparrow, 0}$) at $\omega = \omega_0 - \omega_z$,
the blue sideband ($\Ket{\downarrow, 0} \leftrightarrow \Ket{\uparrow, 1}$)
at $\omega = \omega_0 + \omega_z$, and the carrier transition
($\Ket{\downarrow, 0} \leftrightarrow \Ket{\uparrow, 0}$) at $\omega =
\omega_0$.
Here, $\omega$ and $\phi$ denote the laser frequency and phase,
respectively, and $\Omega$ is the Rabi frequency.
The availability of these three transitions enables the implementation of
arbitrary single- and two-qubit gates, such as the Hadamard gate $\hat{H}
= (\hat{\sigma}_x + \hat{\sigma}_z) / \sqrt{2}$, thereby providing a
complete toolkit for quantum information processing in the spin-motional
Hilbert space.

We initialize the system in the state $\Ket{{\rm init}} =
\Ket{\uparrow, 0}$ prior to the measurement protocol described in the
following subsection.
This state is prepared by means of two laser-driven processes: optical
pumping, which transfers the spin state from $\Ket{\downarrow}$ to
$\Ket{\uparrow}$, and the red sideband transition, $\Ket{\uparrow, n + 1}
\rightarrow \Ket{\downarrow, n}$, which cools the motional mode close to
the ground state.
It has been demonstrated that an initial mean phonon occupation of
$\bar{n}_0 \sim 10^{-3}$ is achievable~\cite{schmidt2000ground}.
State-selective readout is performed via fluorescence detection on the
Doppler cooling transition, for example through site-resolved imaging
with a camera~\cite{Pogorelov:2021eha}.
Fluorescence photons are observed when the ion occupies the state
$\Ket{\downarrow}$, whereas no fluorescence signal is detected for
$\Ket{\uparrow}$.
If fluorescence photons emitted from an ion in $\Ket{\downarrow}$ are not
detected, the state may be misidentified as $\Ket{\uparrow}$.
The resulting readout error has been reported to be $\epsilon_{\rm M}
\sim 10^{-4}$~\cite{Harty:2014tkj}.


\subsection{Search protocol}
\label{sec:protocol}

In this subsection, we outline a detection scheme for weak signals of
the form
\begin{equation}
    \hat{H}_{\rm int} = \alpha \, \hat{a} + \alpha^* \, \hat{a}^\dagger
    \label{eq:Hint}
\end{equation}
in the interaction picture, where $\hat{a}^{(\dagger)}$ denote the ladder
operators of the vibrational qubit.
The corresponding time evolution is described by the displacement operator
\begin{equation}
    \hat{D} (t) \simeq \left( 1 + i \alpha_{\rm i} t \,
    \hat{\sigma}^2_{\rm vib} \right) e^{-i \alpha_{\rm r} t \,
    \hat{\sigma}^1_{\rm vib}} ,
\end{equation}
where $\alpha_{\rm r} = \vert \alpha \vert \cos \phi$ and $\alpha_{\rm i}
= \vert \alpha \vert \sin \phi$ denote the real and imaginary parts of
$\alpha$, respectively, and we define $\hat{\sigma}^1_{\rm vib} =
\hat{a}^\dagger + \hat{a}$ and $\hat{\sigma}^2_{\rm vib} = i
\hat{a}^\dagger - i \hat{a}$.
Here, we assume $\alpha$ to be small and neglect the higher-order terms.

The simplest strategy for detecting such an interaction is to measure
the $\Ket{0} \to \Ket{1}$ transition probability by preparing the
$\Ket{0}$ state, in which case the transition probability is proportional
to $\vert \alpha \vert^2$.
In this work, we instead consider an interferometric approach in which
the signal response becomes linear in $\alpha$
by preparing a different initial state.
The advantage of the $\Ket{0} \to \Ket{1}$ transition strategy lies in
the cleanliness of the signal, whose background consists primarily of
heating noise.
The interferometric approach, on the other hand, requires
statistical inference to extract the signal.
Which type of experiment yields better sensitivity depends on both the
noise rate and the statistical uncertainties of the measurements.

We first prepare the superposition of motional eigenstates $\Ket{+} =
(\Ket{0} + \Ket{1}) / \sqrt{2}$, which is realized by applying the
Hadamard gate $\hat H$ to the spin state, followed by the operation that
maps the spin state onto the motional state via the
sideband laser, $\hat U_{\rm SB} = \exp \big[ (\pi / 2)
(\hat{\sigma}_- \hat{a}^\dagger - \hat{\sigma}_+ \hat{a}) \big]$,
that is, $\Ket{+} = \hat{U}_{\rm SB} \hat{H} \Ket{{\rm init}}$.
Under the interaction Hamiltonian~\eqref{eq:Hint}, the state evolves as
\begin{equation}
    \hat{D} (T) \Ket{+} \simeq e^{-i \alpha_{\rm r} T}
    \left( \frac{1 + \alpha_{\rm i} T}{\sqrt{2}} \Ket{0} +
    \frac{1 - \alpha_{\rm i} T}{\sqrt{2}} \Ket{1} \right)
\end{equation}
after an interrogation time $T$.
The population of $\Ket{2}$ is of order $\alpha^2$ and
is therefore negligible.
After applying $\hat{U}_{\rm SB}^\dagger$ to map the state back onto the
spin basis, the final state of the protocol is
\begin{equation}
    \Ket{{\rm fin}} =
    \frac{1 - \alpha_{\rm i} T}{\sqrt{2}} \Ket{\uparrow} +
    \frac{1 + \alpha_{\rm i} T}{\sqrt{2}} \Ket{\downarrow}
\end{equation}
to the leading order in $\alpha$, where the overall phase
has been omitted.

We prepare $N$ qubits simultaneously and record the
number of outcomes $\Ket{\uparrow}$ and $\Ket{\downarrow}$, denoted
$N_\uparrow$ and $N_\downarrow$, respectively, which follow a binomial
distribution\footnote{Here, all the qubits are assumed to be held in 
separate traps, for simplicity.
In the case where $n$ qubits are in a single trap, one can use
the center-of-mass mode as the single vibrational qubit.
The coupling $\alpha$ is, in this case, effectively enhanced by $\sqrt{n}$
while the number of qubits is reduced to $N/n$.
The scaling of the sensitivity is thus the same
as in the decoupled case, since the SNR scales as $N \alpha^2$.}.
The point is that due to the long-range coherence of the dark-matter
signals, the rotation of the spin direction is common for $N$ qubits even
though the sign of $\alpha_{\rm i}$ is random at each measurement.
The signal is therefore a variation of $|N_\uparrow - N_\downarrow|$
larger than the statistical expectation.
Defining the detection probability for $\Ket{\uparrow}$ as $p = (1 +
\Delta p) / 2$ with $\Delta p = -2 \alpha_{\rm i} T = -2 \vert \alpha
\vert T \sin \phi$, the expectation value and variance of $\Delta N =
N_\uparrow - N_\downarrow$ are given by $\braket{\Delta N} = N \Delta p$
and ${\rm var} (\Delta N) = N$ to leading order in $\Delta p$, respectively.
The absolute value of $\Delta N$ follows a folded normal distribution for
large $N$, whose expectation value and variance are given by
$\braket{\vert \Delta N \vert} = \sqrt{2 N/\pi} (1 + \lambda^2 / 2) +
{\cal O} (\lambda^4)$ and ${\rm var} (\vert \Delta N \vert)_{\lambda = 0}
= N (1 - 2 / \pi)$, respectively, with $\lambda = \sqrt{N} \Delta p$.
Thus, the single-shot signal-to-noise ratio depends only on $\lambda$:
\begin{equation}
    {\rm SNR}_1 = \frac{\braket{\vert \Delta N \vert} -
    \braket{\vert \Delta N \vert}_{\lambda = 0}}{\sqrt{{\rm var} (\vert
    \Delta N \vert)_{\lambda = 0}}} = \frac{\lambda^2}{\sqrt{2 (\pi - 2)}}
    + {\cal O} (\lambda^4) .
\end{equation}

For dark matter detection, the phase $\phi$ is randomly distributed.
Therefore, we adopt the strategy of repeating the measurement 
many times, averaging the expectation value over $\phi$ as
$\braket{\vert \Delta N \vert} = \sqrt{2 N / \pi} (1 + \bar{\lambda}^2 /
4) + {\cal O} (\bar{\lambda}^4)$ with $\bar{\lambda} = {\rm max} (\lambda)
= 2 \sqrt{N} \vert \alpha \vert T$.
The signal-to-noise ratio is modified to
\begin{equation}
    {\rm SNR}_1 = \frac{\braket{\vert \Delta N \vert} -
    \braket{\vert \Delta N \vert}_{\bar{\lambda} = 0}}{\sqrt{{\rm var}
    (\vert \Delta N \vert)_{\bar{\lambda} = 0}}} =
    \frac{\bar{\lambda}^2}{\sqrt{8 (\pi - 2)}}
    + {\cal O} (\bar{\lambda}^4) ,
    \label{eq:snr}
\end{equation}
which is again a function of a single parameter $\bar{\lambda}$.
Repeating this measurement $N_{\rm rep}$ times increases the
signal-to-noise ratio by a factor of $N_{\rm rep}^{1/2}$, i.e.,
${\rm SNR}_{N_{\rm rep}} = N_{\rm rep}^{1/2} \, {\rm SNR}_1$.
The sensitivity to the signal parameter $\eta \equiv \vert \alpha \vert$,
defined through the signal-to-noise ratio as ${\rm SNR}_{N_{\rm rep}} =
1$, is therefore given by
\begin{equation}
    \delta \eta \simeq \left( \frac{\pi - 2}{2} \right)^{1/4}
    N_{\rm rep}^{-1/4} \, N^{-1/2} \, T^{-1} .
\end{equation}

When systematic noise contributions, namely the initial thermal population,
readout errors, and motional heating, are included, the sensitivity becomes
\begin{equation}
    \delta \eta = \left( \frac{\pi - 2}{2} \right)^{1/4}
    N_{\rm rep}^{-1/4} \, N^{-1/2} \, T^{-1} \left[ 1 + (1 - c) \bar{n}_0
    + \frac{\pi}{4 (\pi - 2)} \epsilon_{\rm M} +
    \frac{c^2 \pi}{2 (\pi - 2)} \dot{\bar{n}} T \right]
    \label{eq:sensitivity_eta}
\end{equation}
with $c = \cos (\pi / \sqrt{2})$.
Here, $\dot{\bar{n}}$ denotes the heating rate, which is independently
measured using the same protocol without the Hadamard gate.
The mean offset due to the noise sources $\epsilon_{\rm M}$ and
$\dot{\bar{n}}$ can be subtracted using calibration measurements.
We assume that the residual uncertainty in this subtraction lies below the
projection noise, corresponding to the systematic errors of
$\epsilon_{\rm M}$ and $\dot{\bar{n}} T$ that are much smaller than
$1/\sqrt{N}$.
The derivation of this formula is presented in Appendix~\ref{sec:error}.
When the calibration of $\epsilon_{\rm M}$ and $\dot{\bar{n}}$ is not
sufficiently precise, that is, the errors of the noise parameters,
denoted as $\sigma_{\rm bkg}$, are above $1/\sqrt{N}$, the scaling is
modified to
\begin{equation}
    \delta \eta = \left( \frac{\pi - 2}{2} \right)^{1/4}
    N_{\rm rep}^{-1/4} \, N^{-1/2} \, T^{-1} \left[ 1 + (1 - c)
    \bar{n}_0 \right] \left[ 1 + \frac{\pi N}{\pi - 2} \sigma_{\rm bkg}^2
    \right]^{1/4} .
\end{equation}

Even if the trap potential is controlled with high precision,
fluctuations in the motional frequency induced by thermal noise
remain~\cite{Brownnutt:2014ifp}.
If $\omega_z$ fluctuates at the kHz level around $\omega_z \sim$ MHz,
i.e., $\delta \omega_z / \omega_z \sim 10^{-3}$ the interrogation
time is effectively limited by its inverse bandwidth, i.e., $T \sim 1$~ms.
This limitation is common to the Rabi-type measurements but has not been
taken into account in previous works~\cite{Ito:2023zhp,Takai:2025cyy}.

Since the sensitivity of the Rabi-type protocol is determined by the
signal-to-noise ratio $S / \sqrt{B} = 1$ with $S = N_{\rm rep} N \delta
\eta^2 T^2$ and $B = N_{\rm rep} N (\epsilon_{\rm M} + \dot{\bar{n}}
T)$~\cite{Ito:2023zhp}, the Ramsey-type protocol becomes advantageous in
the regime $K = N (\epsilon_{\rm M} + \dot{\bar{n}} T) \gg 1$,
in which the sensitivity is improved by a factor of $K^{1/4}$.
If the calibration error $\sigma_{\rm bkg}$ is larger than $1 / \sqrt{N}$,
the improvement factor from the Rabi-type protocol does not scale
as $N$ and is given by the relative error of the noise parameters.


\section{Sensitivity to wave-like dark matter}

We consider wave-like background fields that induce weak, coherent
electric fields within linear Paul traps and can resonantly excite
their collective motional modes.
For wave-like dark matter with mass $m_{\rm DM}$, the field takes the
form $\Phi = \Phi_0 \cos (m_{\rm DM} t - \phi)$, where $\Phi_0$ is the
amplitude and $\phi$ is a random phase that remains approximately
constant over the coherence time $T_{\rm DM} = 2 \pi / m_{\rm DM}
v_{\rm DM}^2$, with $v_{\rm DM} \sim 10^{-3}$ denoting the relative
velocity of the dark matter.
The amplitude is determined by the local dark matter energy density
$\rho_{\rm DM}$ through $\Phi_0 = \sqrt{2 \rho_{\rm DM}} / m_{\rm DM}$.
We consider two benchmark scenarios, the axion-like particle and the
dark photon, both of which induce an effective classical electric field
capable of resonantly driving the center-of-mass mode.
Throughout this work, we refer to the axion-like particle simply as the
axion.

The axion field $a$ induces an electric field via ${\cal L} = -
(g_{a \gamma} / 4) a F_{\mu\nu} \tilde{F}^{\mu\nu}$, where $F_{\mu\nu}$
is the electromagnetic field strength tensor, and $\tilde{F}_{\mu\nu}$
its dual.
In the presence of a background magnetic field $B$ oriented along the
$z$-axis, the axion-induced electric field in the regime $m_{\rm DM} R
\ll 1$ is given by $E_z \sim (g_{a\gamma} / m_{\rm DM}) B (m_{\rm DM} R)^2
\sqrt{2 \rho_{\rm DM}} \sin (m_{\rm DM} t - \phi)$, where $R$ denotes
the characteristic size of the magnetic field~\cite{Ito:2023zhp}.
The ${\cal O} (1)$ prefactor depends on the geometry of both the magnetic
field and the cavity.
At resonance, $m_{\rm DM} = \omega_z$, the interaction
Hamiltonian takes the form of Eq.~\eqref{eq:Hint} with
\begin{equation}
    \alpha \sim \frac{g_{a \gamma} e B}{2 m_{\rm DM}} (m_{\rm DM} R)^2
    \sqrt{\frac{\rho_{\rm DM}}{m_{\rm ion} \omega_z}} \, e^{i \phi} ,
\end{equation}
where $e$ is the elementary charge and $\phi$ absorbs all
complex phases.

In the dark photon scenario, the dark photon field $A'$ couples to the
Standard Model photon through kinetic mixing, ${\cal L} = (\epsilon / 2)
F^{\mu\nu} F'_{\mu\nu}$, where $F'_{\mu\nu} = \partial_\mu A'_\nu -
\partial_\nu A'_\mu$ is the dark photon field strength tensor.
The induced effective electric field along the $z$-axis is $E_z =
\epsilon \cos \theta \sqrt{2 \rho_{\rm DM}} \sin (m_{\rm DM} t - \phi)$,
where the polarization angle $\theta$ is assumed to be isotropically
distributed~\cite{Ito:2023zhp}.
Here, we assume that no boundary conditions are imposed on the electric
field.
At resonance, $m_{\rm DM} = \omega_z$, the interaction Hamiltonian is
given by Eq.~\eqref{eq:Hint} with
\begin{equation}
    \alpha = \frac{e \epsilon}{2} \sqrt{\frac{\rho_{\rm DM}}{m_{\rm ion}
    \omega_z}} \cos \theta \, e^{i \phi} ,
\end{equation}
where all complex phases are absorbed into $\phi$.

Hereinafter, we present the projected sensitivities to wave-like
dark matter.
(Detection prospects for high-frequency gravitational waves are
discussed in Appendix~\ref{sec:gw}.)
We assume that the three noise sources appearing in
Eq.~\eqref{eq:sensitivity_eta} are sufficiently small to be neglected.
The sensitivity is given by
\begin{equation}
    \delta \eta = 1.6~\text{mHz} \times \left( \frac{T}{10~{\rm ms}}
    \right)^{-3/4} \left( \frac{N}{10^6} \right)^{-1/2}
    \left( \frac{T_{\rm tot}}{1~{\rm day}} \right)^{-1/4} ,
\end{equation}
where the total observation time is defined as $T_{\rm tot} =
N_{\rm rep} T$.

\begin{figure}[t!]
    \centering
    \includegraphics[width=0.8\linewidth]{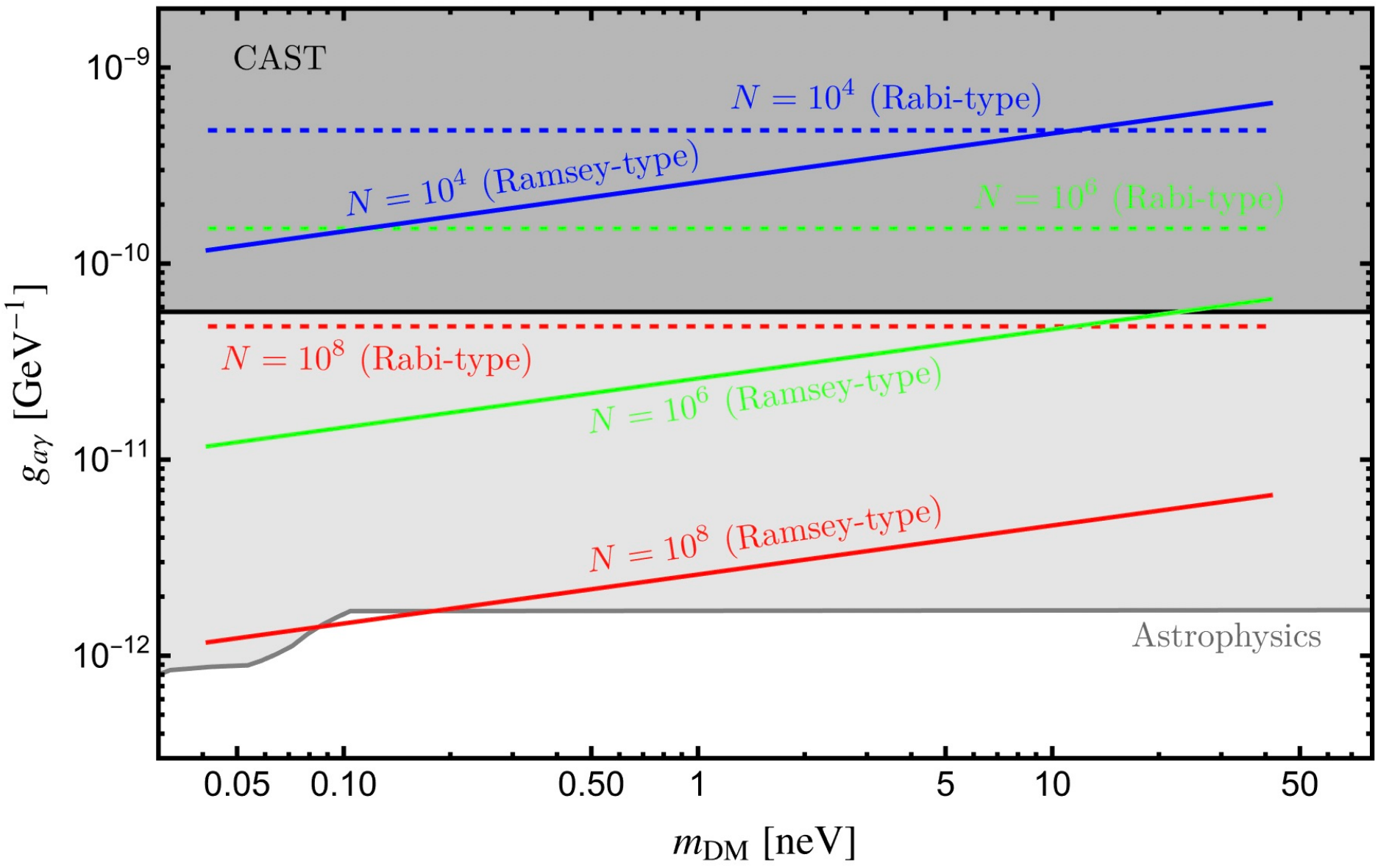}
    \caption{Sensitivity to the axion-photon coupling $g_{a \gamma}$
    for $N = 10^4$ (blue), $10^6$ (green), and $10^8$ (red) ions.
    The solid and dashed lines correspond to the Ramsey-type protocol
    proposed in this work and the Rabi-type protocol of
    Ref.~\cite{Ito:2023zhp}, respectively.
    The sensitivities are computed assuming a one-year observation
    period for scanning the axion dark matter mass range,
    $10~{\rm kHz} \leq m_{\rm DM} / 2 \pi \leq 10~{\rm MHz}$.
    The black and gray shaded regions are excluded by
    CAST~\cite{CAST:2024eil} and astrophysical
    observations~\cite{Ning:2024eky,Benabou:2025jcv}, respectively.}
    \label{fig:axion}
\end{figure}

For axion dark matter, the sensitivity to the axion-photon coupling is
obtained as
\begin{align}
    g_{a \gamma} &= 7.0 \times 10^{-12}~{\rm GeV}^{-1} \times \left(
    \frac{N}{10^6} \right)^{-1/2} \left( \frac{m_{\rm DM} T}{10^4}
    \right)^{-3/4} \left( \frac{T_{\rm tot}}{1~{\rm day}} \right)^{-1/4}
     \left( \frac{m_{\rm DM}}{1~{\rm neV}} \right)^{1/4} \notag \\
    &\qquad \qquad \times \left( \frac{m_{\rm ion}}{37~{\rm GeV}}
    \right)^{1/2} \left( \frac{R}{3~{\rm m}} \right)^{-2}
    \left( \frac{B}{100~{\rm mT}} \right)^{-1} \left(
    \frac{\rho_{\rm DM}}{0.45~{\rm GeV}~{\rm cm}^{-3}} \right)^{-1/2} ,
    \label{eq:axion}
\end{align}
where we average over the phase $\phi$.
Fig.~\ref{fig:axion} shows the projected sensitivity to
$g_{a \gamma}$ for $N = 10^4$ (blue), $10^6$ (green), and $10^8$ (red)
ions, assuming that the entire mass range $10~{\rm kHz} \leq m_{\rm DM} /
2 \pi \leq 10~{\rm MHz}$ is scanned over a total observation time of
one year.
Here, the width of the mass bin at each dark matter mass $m_{\rm DM}$
is set to $10^{-4} m_{\rm DM}$, so that a time of $\simeq 7.6$~min
is spent on each bin, supposing that the fluctuation of trap frequency is
order of 100~Hz.
The Ramsey-type protocol is shown as the solid lines with the experimental
parameters set to those in Eq.~\eqref{eq:axion} except for $T_{\rm tot}$.
For comparison, the dashed lines show the Rabi-type
protocol~\cite{Ito:2023zhp} using the same parameters with $\dot{\bar{n}}
= 0.1$~s$^{-1}$ and $\epsilon_{\rm M} \ll \dot{\bar{n}} T$.
The gray shaded region represents astrophysical
bounds~\cite{Ning:2024eky,Benabou:2025jcv}, while the black shaded region
is excluded by CAST~\cite{CAST:2024eil}.
The data points are taken from Ref.~\cite{AxionLimits}.
For $N = 10^6$, the projected sensitivity can surpass the CAST bound,
while for $N = 10^8$, it can be comparable to
the current astrophysical bounds.
Note that all ions need not be confined within a single trap.
The requirement is that all ions remain within the coherence length, i.e.,
${\cal O} (100)$ kilometers, and cooled to the motional ground state.
No fundamental technological obstacle is anticipated in achieving large
$N$, although a large-scale array of ion traps would be required to reach,
for example, $N = 10^8$.
This number can be compared with the number of qubits required for future
practically useful quantum computers, which is estimated to be at least
$\sim 10^6$.

\begin{figure}[t!]
    \centering
    \includegraphics[width=0.8\linewidth]{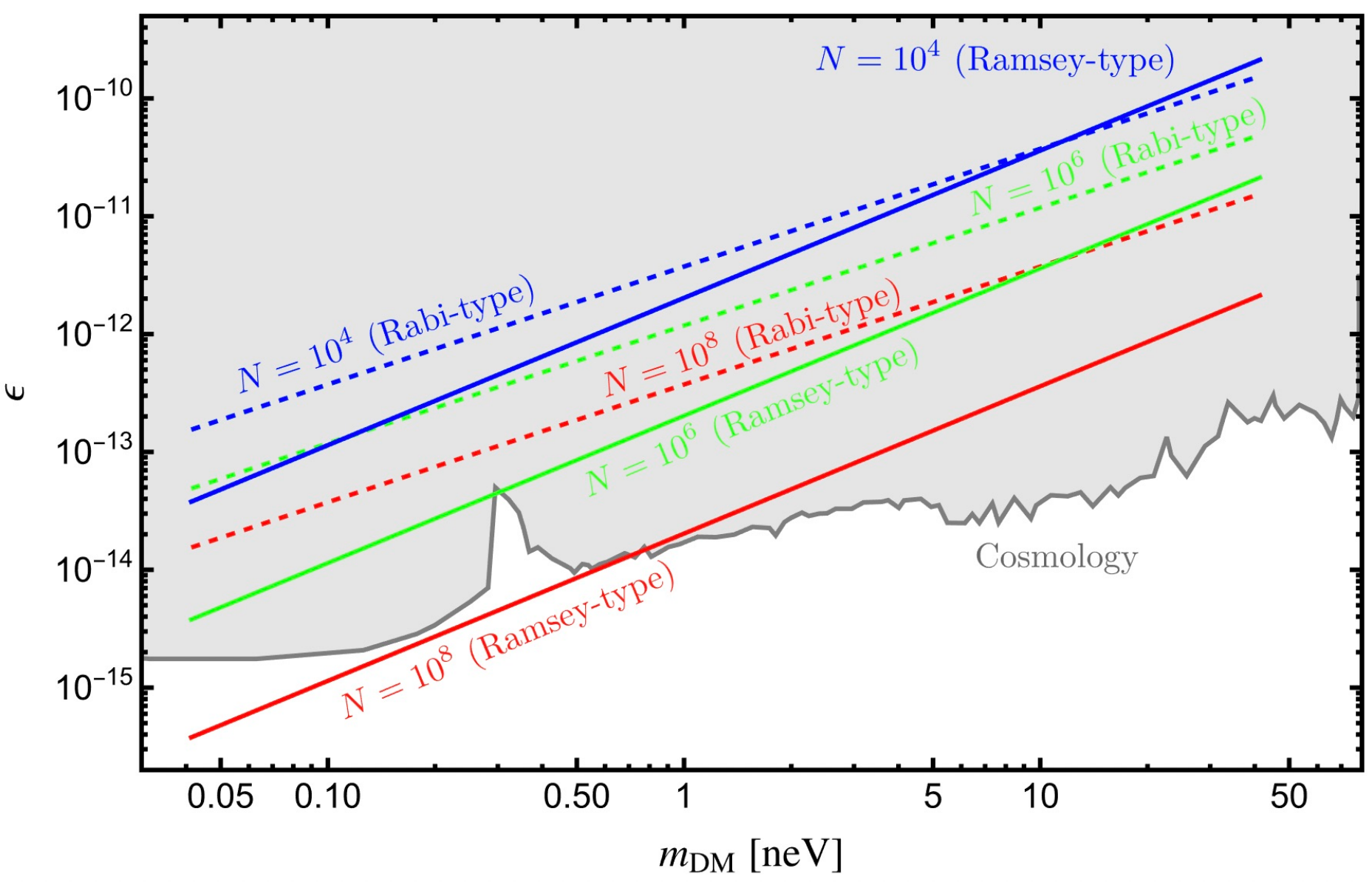}
    \caption{Sensitivity to the kinetic mixing parameter of the dark
    photon $\epsilon$ for $N = 10^4$ (blue), $10^6$ (green), and
    $10^8$ (red) ions.
    The solid and dashed lines correspond to the Ramsey-type protocol
    proposed in this work and the Rabi-type protocol of
    Ref.~\cite{Ito:2023zhp}, respectively.
    The sensitivities are computed assuming a one-year observation period
    for scanning the dark photon dark matter mass range, $10~{\rm kHz}
    \leq m_{\rm DM} / 2 \pi \leq 10~{\rm MHz}$.
    The gray shaded region is excluded by cosmological
    observations~\cite{Arias:2012az,Witte:2020rvb,An:2024wmc}.}
    \label{fig:darkphoton}
\end{figure}

For dark photon dark matter, the sensitivity to the kinetic mixing
parameter is obtained as
\begin{align}
    \epsilon &= 5.5 \times 10^{-14} \times \left( \frac{N}{10^6}
    \right)^{-1/2} \left( \frac{m_{\rm DM} T}{10^4} \right)^{-3/4}
    \left( \frac{T_{\rm tot}}{1~{\rm day}} \right)^{-1/4} \notag \\
    &\qquad \qquad \qquad \times \left( \frac{m_{\rm ion}}{37~{\rm GeV}}
    \right)^{1/2} \left( \frac{m_{\rm DM}}{1~{\rm neV}} \right)^{5/4}
    \left( \frac{\rho_{\rm DM}}{0.45~{\rm GeV}~{\rm cm}^{-3}}
    \right)^{-1/2} ,
\end{align}
where we average over the polarization angle $\theta$ and
the phase $\phi$.
Fig.~\ref{fig:darkphoton} shows the projected sensitivity to $\epsilon$
for $N = 10^4$ (blue), $10^6$ (green), and $10^8$ (red) ions, assuming
that the full mass range $10~{\rm kHz} \leq m_{\rm DM} / 2 \pi \leq
10~{\rm MHz}$ is scanned within a total observation time of one year.
We again take the bin width to be $10^{-4} m_{\rm DM}$.
The gray shaded region corresponds to cosmological
bounds~\cite{Arias:2012az,Witte:2020rvb,An:2024wmc}.
The data points are taken from Ref.~\cite{AxionLimits}.
For $N = 10^6 \text{--} 10^8$, the projected sensitivity can surpass the
existing cosmological bounds in the sub-neV mass range.

\section{Conclusions}

We have investigated a new quantum protocol for wave-like dark matter
detection, based on an array of qubits.
If dark matter interacts with the qubits, it can slightly change the
state of the qubits from their initially prepared state.
The signal of dark matter is this small modification.
One can consider the Rabi-type protocol, in which a qubit is prepared in
$\Ket{0}$ and a dark-matter-induced transition to $\Ket{1}$ is sought,
whose probability is proportional to $\alpha^2$, where $\alpha$ is the
coupling of dark matter to the qubit.
With $N$ detectors, this Rabi-type scheme gives a signal-to-noise ratio
of ${\rm SNR} \sim \sqrt{N} \alpha^2$.
Quantum resources such as entanglement can improve this scaling as
${\rm SNR} \sim N \alpha^2$ or $N^{3/2} \alpha^2$, but they remain
challenging to implement with current technology for large $N$.

We have proposed a Ramsey-type measurement protocol based on a
superposition of the qubit states, $\Ket{0} + \Ket{1}$.
Since dark matter acts coherently on an array of qubits within its
coherence length, the numbers of 0 and 1 outcomes can be
compared in a single measurement, leaving an imbalance proportional to
$N^{3/2} \alpha^2$.
The resulting signal-to-noise ratio is ${\rm SNR} \sim N \alpha^2$,
so the sensitivity to the coupling scales as $\delta \alpha \sim
N^{-1/2}$, the same as that of the maximally entangled scheme in a
pessimistic scenario.
Since the protocol involves only simple quantum operations on each qubit,
we foresee no fundamental obstacle to scaling it up, although a
large-scale array of ion traps would be required.

We have demonstrated the potential of this new protocol, employing
linear Paul traps as an example platform.
We have focused on axion and dark photon dark matter in the neV mass
range, both of which generate effective classical electric fields capable
of resonantly driving the collective motional modes of trapped ions.
We have derived the sensitivity including systematic noise contributions
from the initial thermal population, readout errors, and motional heating
and have found that the statistical uncertainty dominates over these
systematic errors for sufficiently precise ion traps.
On the other hand, the Rabi-type protocol is limited mainly by heating
due to environmental photons.
The Ramsey-type protocol proposed here outperforms the Rabi-type approach
in the regime $N \dot{\bar{n}} T \gg 1$, with $\dot{\bar{n}}$ and $T$
denoting the heating rate and the single-shot measurement time, respectively.

We have illustrated the projected sensitivities to axion and dark photon
dark matter for $N = 10^4$, $10^6$, and $10^8$ ions in the mass range
$10~{\rm kHz} \leq m_{\rm DM} / 2 \pi \leq 10~{\rm MHz}$, together with
those of the Rabi-type protocol and current astrophysical and
cosmological bounds.
Notably, for the experimental parameters considered in this work, the
Rabi-type protocol would require $N = 10^{13} \text{--} 10^{14}$ ions to
reach the existing astrophysical constraints, whereas the Ramsey-type
measurement achieves comparable sensitivity with only $N = 10^6$--$10^8$
ions.
The same platform can also search for high-frequency gravitational waves,
for which a similar protocol applies.

We would like to emphasize that our protocol is not specific to a
particular platform and can be applied to other qubit systems.
The difference would be the frequency ranges 
for each platform, especially the necessity of 
the synchronization of qubit operations.
For the ion-trap implementation considered here, the relevant motional frequencies lie in the kHz--MHz range, so the synchronization requirement is much less stringent than in platforms operating at GHz transition frequencies.
Exploring similar protocols in other qubit platforms would be an interesting direction for future research.


\section*{Acknowledgment}

We would like to thank Utako Tanaka for
useful discussions as well as for
providing us with information on the
latest developments in ion-trap devices.
We would also like to thank Hajime Fukuda, Roni Harnik, and Zachary Bogorad 
for useful discussions and comments on $N$ scaling of the measurements.
This work is supported by JSPS KAKENHI Grant Numbers 22K21350~(RK) and 24KJ1157~(RT).


\appendix
\section{Sensitivity including systematic errors}
\label{sec:error}

In this appendix, we derive the sensitivity
formula~\eqref{eq:sensitivity_eta}.
The initial state takes the form
\begin{equation}
    \rho_0 = \Ket{\uparrow} \Bra{\uparrow} \otimes \sum_n p_n (\bar{n}_0)
    \Ket{n} \Bra{n} ,
\end{equation}
where the vectors $\Ket{n}$ satisfy $\hat{a}^\dagger \hat{a} \Ket{n} =
n \Ket{n}$.
The function $p_n (\bar{n}_0)$ represents thermal occupation probabilities
of the initial motional states with $\bar{n}_0 = \sum_n n p_n (\bar{n}_0)$.
After applying the Hadamard gate and the sideband transition, the state
evolves into
\begin{equation}
    \rho_1 = \hat{U}_{\rm SB} \hat{H} \rho_0 \hat{H}^\dagger
    \hat{U}_{\rm SB}^\dagger = \sum_n p_n (\bar{n}_0) \Ket{\psi (n)}
    \Bra{\psi (n)}
\end{equation}
with
\begin{equation}
    \begin{split}
        \Ket{\psi (n)} &= \Ket{\uparrow} \otimes \frac{1}{\sqrt{2}}
        \left[ \cos \frac{\pi \sqrt{n+1}}{2} \Ket{n} - \sin
        \frac{\pi \sqrt{n}}{2} \Ket{n-1} \right] \\
        &\quad + \Ket{\downarrow} \otimes \frac{1}{\sqrt{2}} \left[ \cos
        \frac{\pi \sqrt{n}}{2} \Ket{n} + \sin \frac{\pi \sqrt{n+1}}{2}
        \Ket{n+1} \right] .
    \end{split}
\end{equation}

It has been demonstrated that the initial phonon occupation $\bar{n}_0 \sim
10^{-3}$ is achievable~\cite{schmidt2000ground}, and we therefore ignore the
higher-order terms in $\bar{n}_0$.
In this approximation, only the lowest two phonon states, $n = 0$
and $n = 1$, are relevant.
In particular, we find $p_0 (\bar{n}_0) = 1 - \bar{n}_0$,
$p_1 (\bar{n}_0) = \bar{n}_0$,
\begin{equation}
    \Ket{\psi (0)} = \Ket{\downarrow} \otimes \left[ \frac{1}{\sqrt{2}}
    \Ket{0} + \frac{1}{\sqrt{2}} \Ket{1} \right] , \quad
    \Ket{\psi (1)} = \Ket{\uparrow} \otimes \left[ -\frac{1}{\sqrt{2}}
    \Ket{0} + \frac{c}{\sqrt{2}} \Ket{1} \right] +
    \Ket{\downarrow} \otimes \frac{s}{\sqrt{2}} \Ket{2}
\end{equation}
with $s = \sin (\pi / \sqrt{2}) \simeq 0.796$ and $c = \cos (\pi /
\sqrt{2}) \simeq -0.606$, and thus
\begin{equation}
    \rho_1 = \sum_{\kappa, \kappa'} \ket{\kappa} \bra{\kappa'}
    \otimes \rho_{\kappa \kappa'} + {\cal O} \left( \bar{n}_0^2 \right)
\end{equation}
with
\begin{align}
    \rho_{\uparrow \uparrow} &= \frac{\bar{n}_0}{2} \Big[ \Ket{0}
    \Bra{0} - c \Ket{0} \Bra{1} - c \Ket{1} \Bra{0} + c^2
    \Ket{1} \Bra{1} \Big] , \\
    \rho_{\downarrow \uparrow} &= \frac{\bar{n}_0 s}{2} \Big[ - \Ket{2}
    \Bra{0} + c \Ket{2} \Bra{1} \Big] , \\
    \rho_{\uparrow \downarrow} &= \rho_{\downarrow \uparrow}^\dagger , \\
    \rho_{\downarrow \downarrow} &= \frac{1 - \bar{n}_0}{2} \Big[
    \Ket{0} \Bra{0} + \Ket{0} \Bra{1} + \Ket{1} \Bra{0} + \Ket{1} \Bra{1}
    \Big] + \frac{\bar{n}_0 s^2}{2} \Ket{2} \Bra{2} .
\end{align}

We then evaluate the expectation value of  $\hat{\sigma}_z$, which is
equivalent to measuring the operator
\begin{equation}
    \hat{M} = \hat{U}_{\rm SB} \, \hat{\sigma}_z \,
    \hat{U}_{\rm SB}^\dagger
\end{equation}
before applying the inverse sideband operation.
Note that $\hat{M}^2 = 1$.
The matrix elements of $\hat{M}$ in the spin basis,
$\hat{M}_{\kappa \kappa'} \equiv \Bra{\kappa} \hat{M} \Ket{\kappa'}$,
are given by
\begin{equation}
    \hat{M}_{\uparrow \uparrow} = \hat{C}^2_1 - \hat{S}^\dagger \hat{S} ,
    \quad \hat{M}_{\downarrow \uparrow} =
    \hat{M}_{\uparrow \downarrow}^\dagger = \hat{S} \hat{C}_1 + \hat{C}_0
    \hat{S} , \quad \hat{M}_{\downarrow \downarrow} = \hat{S}
    \hat{S}^\dagger - \hat{C}^2_0
\end{equation}
with
\begin{equation}
    \hat{C}_\lambda = \sum_{k=0}^\infty \frac{(-1)^k}{(2k)!} \left(
    \frac{\pi}{2} \right)^{2k} (\hat{a}^\dagger \hat{a} + \lambda)^k,
    \quad \hat{S} = \sum_{k=0}^\infty \frac{(-1)^k}{(2k+1)!} \left(
    \frac{\pi}{2} \right)^{2k+1} (\hat{a}^\dagger \hat{a})^k \hat{a}^\dagger
\end{equation}
for $\lambda = 0$, 1.
The expectation value of the signal observable, $\Delta p =
\Braket{\hat{\sigma}_z}$ is calculated as
\begin{equation}
    \Delta p = \sum_{\kappa, \, \kappa'} {\rm Tr} \Big[
    \hat{M}_{\kappa \kappa'} \, \rho_{\kappa' \kappa} (T) \Big] ,
    \label{eq:ev}
\end{equation}
where $\rho_{\kappa \kappa'} (T)$ denotes the density matrix
$\rho_{\kappa \kappa'}$ evolved under the target field during the
measurement time $T$ in the interaction picture, and the trace is
taken over the vibrational degrees of freedom.

The master equation in the interaction picture is given
by~\cite{Henkel:1999eh,Brownnutt:2014ifp}
\begin{equation}
    \dot{\rho} = i \left[ \rho, \hat{H}_{\rm int} (t) \right] + \gamma
    (\bar{N} + 1) \left( \hat{a} \rho \hat{a}^\dagger - \frac{1}{2}
    \left\{ \rho, \hat{a}^\dagger \hat{a} \right\} \right) + \gamma
    \bar{N} \left( \hat{a}^\dagger \rho \hat{a} - \frac{1}{2} \left\{
    \rho, \hat{a} \hat{a}^\dagger \right\} \right) ,
\end{equation}
where $\{ X, Y \} = X Y + Y X$ denotes the anticommutator.
To linear order in $\alpha$ and $\gamma \bar{N} \simeq
\dot{\bar{n}}$, we obtain
$\Delta p = (\Delta p)_{\rm bkg} + (\Delta p)_{\rm sig}$,
\begin{equation}
    (\Delta p)_{\rm bkg} = \epsilon_{\rm M} + 2 c^2 \dot{\bar{n}} T, \quad
    (\Delta p)_{\rm sig} = - 2 b \eta T \sin \phi
\end{equation}
with $c = \cos (\pi / \sqrt{2})$, $b = 1 - (1-c) \bar{n}_0$, and
${\rm Im} \, \alpha = \eta \sin \phi$.
Here, the readout error $\epsilon_{\rm M}$ is incorporated.

We assume that the calibration measurement of the noise parameters,
$\bar{n}_0$, $\epsilon_{\rm M}$ and $\dot{\bar{n}}$, is performed
concurrently with the Rabi-type search using the same $N$ qubits
and is therefore repeated $N_{\rm rep}$ times along with the search.
The uncertainty of the offset $N (\Delta p)_{\rm bkg}$ is given by
$[N (\Delta p)_{\rm bkg}]^{1/2}$ for a single calibration,
ignoring the systematic error.
One obtains the statistical quantities, e.g., the expectation value and
variance of $\Delta N = N_\uparrow - N_\downarrow - N (\Delta p)_{\rm bkg}$
and its absolute value from the formulae in Sec.~\ref{sec:protocol} with
the replacement $\Delta p \to (\Delta p)_{\rm sig}$.
The signal-to-noise ratio is given by
\begin{equation}
    {\rm SNR}_1 = \frac{\braket{\vert \Delta N \vert} - \braket{\vert
    \Delta N \vert}_{\eta = 0}}{\sqrt{{\rm var} (\vert \Delta N
    \vert)_{\eta = 0} + N (\Delta p)_{\rm bkg}}} = \frac{b^2
    \bar{\lambda}^2}{\sqrt{8 (\pi - 2 + \pi (\Delta p)_{\rm bkg})}} ,
\end{equation}
where the random phase $\phi$ has been averaged out.
When $b = 1$ and $(\Delta p)_{\rm bkg} = 0$, this reduces to the
signal-to-noise ratio without noise, Eq.~\eqref{eq:snr}.
Imposing ${\rm SNR}_{N_{\rm rep}} = N_{\rm rep}^{1/2} \, {\rm SNR}_1 = 1$
to estimate the sensitivity, one obtains
\begin{equation}
    \delta \eta = \left( \frac{\pi - 2}{2} \right)^{1/4}
    N_{\rm rep}^{-1/4} \, N^{-1/2} \, T^{-1} \left[ 1 + (1 - c) \bar{n}_0
    + \frac{\pi}{4 (\pi - 2)} \epsilon_{\rm M} +
    \frac{c^2 \pi}{2 (\pi - 2)} \dot{\bar{n}} T \right]
\end{equation}
up to the linear order of the noise parameters.


\section{Sensitivity to high-frequency gravitational waves}
\label{sec:gw}

We further consider high-frequency gravitational waves described by the
metric perturbation $h_{ij} (t) = h_0 e_{ij} \cos (\omega t - {\bm k}
\cdot {\bm x} - \phi)$, where $h_0$ is the strain amplitude, $\omega = 2
\pi f$ is the angular frequency, ${\bm k}$ is the wavevector, and
\begin{equation}
    e_{ij} = \frac{1}{\sqrt{2}}
    \begin{pmatrix}
        \cos^2 \theta & \cos \theta & - \sin \theta \cos \theta \\
        \cos \theta & -1 & -\sin \theta \\
        - \sin \theta \cos \theta & -\sin \theta & \sin^2 \theta
    \end{pmatrix}
\end{equation}
is the polarization tensor.
For simplicity, we assume the gravitational wave to be monochromatic and
unpolarized.

Gravitational waves couple to the trapped ions through two distinct
mechanisms.
First, in the presence of a background magnetic field, graviton-photon
conversion generates an effective electric field $E_z \sim h_0 B
(\omega R)^2 \sin^2 \theta \sin (\omega t - \phi)$ for $\omega R \ll 1$
in the proper detector frame~\cite{Takai:2025cyy}, which resonantly
drives the center-of-mass mode in direct analogy to the axion case.
At resonance, $\omega = \omega_z$, the interaction Hamiltonian takes
the form of Eq.~\eqref{eq:Hint} with
\begin{equation}
    \alpha \sim \frac{h_0}{2} \frac{e B}{\sqrt{2 m_{\rm ion}
    \omega_z}} (\omega_z R)^2 \sin^2 \theta \, e^{i \phi} ,
\end{equation}
where $\phi$ absorbs all irrelevant complex phases.

Second, gravitational waves couple directly to the spatial configuration
of the ion chain through tidal forces of the form $H_{\rm int} =
\frac{1}{2} m_{\rm ion} R_{0k0l} x^k x^l$, where $R_{\mu\nu\rho\sigma}$
is the Riemann tensor evaluated at the trap center~\cite{Ito:2020xvp}.
In a two-ion configuration confined within a single trap, two normal
modes arise: the center-of-mass mode and the stretch mode, and the
latter enables direct detection of this effect.
The free Hamiltonian reads
\begin{equation}
    \hat{H}_0 = \frac{\omega_0}{2} \hat{\sigma}_z^1 + \frac{\omega_0}{2}
    \hat{\sigma}_z^2 + \omega_z \hat{a}^\dagger_{\rm com} \hat{a}_{\rm com}
    + \omega'_z \hat{a}^\dagger_{\rm str} \hat{a}_{\rm str}
\end{equation}
with $\omega'_z = \sqrt{3} \, \omega_z$.
Here, $\hat{\sigma}_z^i$ denotes the Pauli-$z$ operator acting on the $i$th
spin, and $\hat{a}^{(\dagger)}_{\rm com, \, str}$ represent the ladder
operators of the center-of-mass and stretch modes, respectively.
When the gravitational wave frequency satisfies $\omega = \omega'_z =
\sqrt{3} \, \omega_z$, the stretch mode is resonantly driven, and the
interaction Hamiltonian takes the form of Eq.~\eqref{eq:Hint} with
\begin{equation}
    \alpha = \frac{h_0}{2 \sqrt{2}} \left( \frac{9}{16} \alpha_{\rm EM}
    m_{\rm ion} \omega'^5_z \right)^{1/6} \sin^2 \theta \, e^{i \phi} ,
\end{equation}
where $\hat{a}$ is replaced by $\hat{a}_{\rm str}$~\cite{Takai:2025cyy}.
Here, $\alpha_{\rm EM} = e^2 / 4 \pi$ is the fine-structure constant.
Notably, this second mechanism does not require an external magnetic
field and provides a way to distinguish signals originating from
wave-like dark matter from those of high-frequency gravitational waves.

\begin{figure}[t!]
    \centering
    \includegraphics[width=0.8\linewidth]{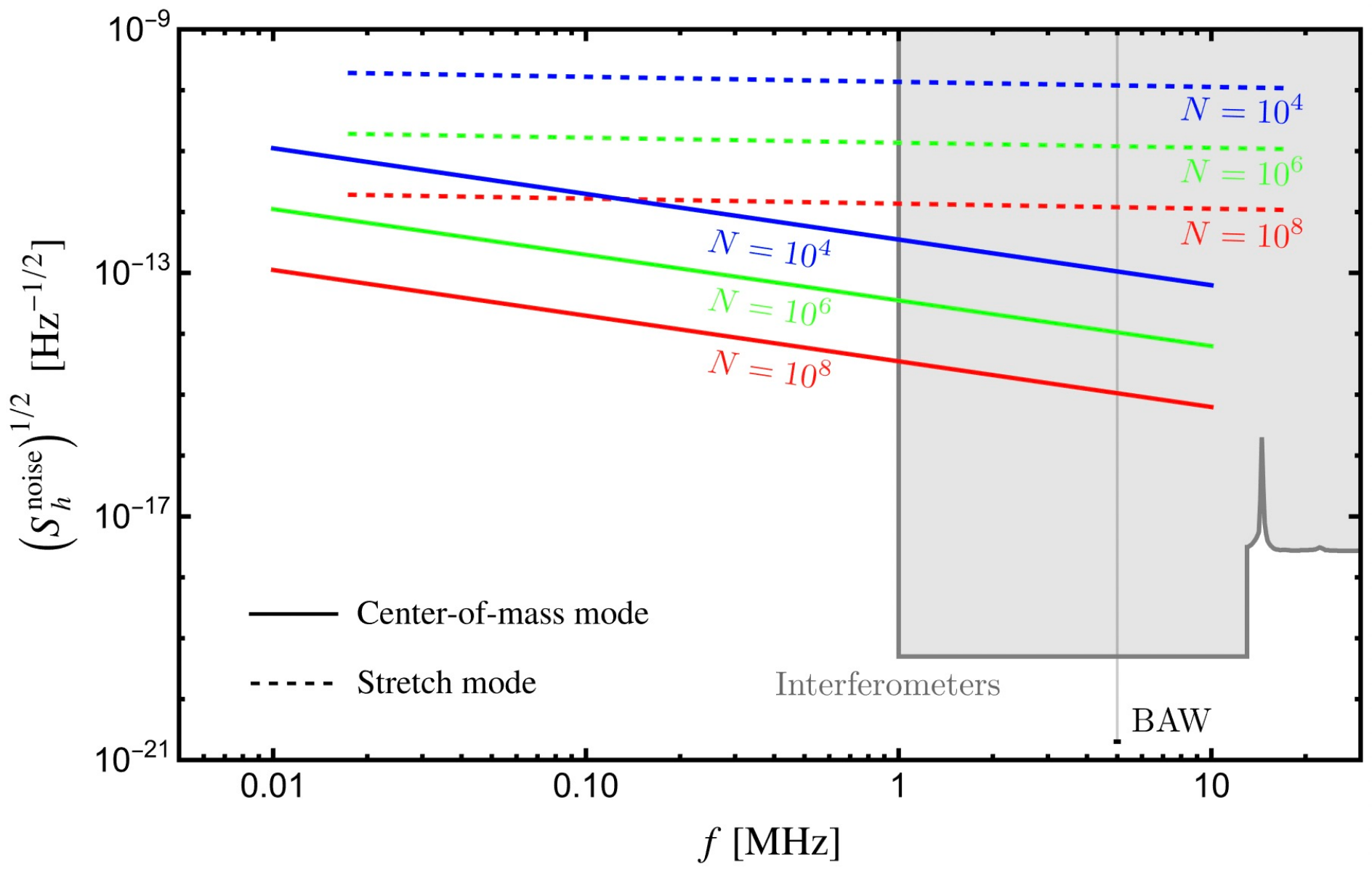}
    \caption{Sensitivity to the noise-equivalent spectral density
    $S_h^{\rm noise} = T h_0^2$ for $N = 10^4$ (blue), $10^6$
    (green), and $10^8$ (red) ions using the center-of-mass mode
    (solid) and the stretch mode (dashed).
    The frequency range $10~{\rm kHz} \leq f \leq 10~{\rm MHz}$ is
    scanned assuming a total observation time of one year.
    The gray and black shaded regions indicate the sensitivities of
    existing experiments based on
    interferometers~\cite{Holometer:2016qoh,Patra:2024eke} and bulk
    acoustic wave (BAW) devices~\cite{Goryachev:2014yra}, respectively.}
    \label{fig:GW}
\end{figure}

The projected sensitivities to the gravitational wave strain $h_0$ through
the center-of-mass and stretch modes are obtained as
\begin{align}
    h_0 &= 8.7 \times 10^{-15} \times \left( \frac{N}{10^6} \right)^{-1/2}
    \left( \frac{\omega T}{10^4} \right)^{-3/4} \left(
    \frac{T_{\rm tot}}{1~{\rm day}} \right)^{-1/4} \notag \\
    &\qquad \qquad \times \left( \frac{m_{\rm ion}}{37~{\rm GeV}}
    \right)^{1/2} \left( \frac{f}{1~{\rm MHz}} \right)^{-3/4} \left(
    \frac{R}{3~{\rm m}} \right)^{-2} \left( \frac{B}{100~{\rm mT}}
    \right)^{-1}
    \label{eq:GWcom}
\end{align}
and
\begin{align}
    h_0 &= 3.4 \times 10^{-12} \times \left( \frac{N}{10^6} \right)^{-1/2}
    \left( \frac{\omega T}{10^4} \right)^{-3/4} \left(
    \frac{T_{\rm tot}}{1~{\rm day}} \right)^{-1/4} \notag \\
    &\qquad \qquad \qquad \qquad \qquad \times \left(
    \frac{m_{\rm ion}}{159~{\rm GeV}} \right)^{-1/6} \left(
    \frac{f}{\sqrt{3} \times 1~{\rm MHz}} \right)^{-1/12} ,
\end{align}
respectively, where we average over the angle $\theta$ and the
phase $\phi$.
Fig.~\ref{fig:GW} shows the projected sensitivity to the noise-equivalent
spectral density, $S_h^{\rm noise} = T h_0^2$, for $N = 10^4$ (blue),
$10^6$ (green), and $10^8$ (red) ions, assuming that the entire frequency
range $10~{\rm kHz} \leq f \leq 10~{\rm MHz}$ is scanned within a total
observation time of one year.
For each frequency $f$, we adopt a frequency bin width of $10^{-4} f$.
The gray and black shaded regions correspond to the sensitivities of
existing experiments based on
interferometers~\cite{Holometer:2016qoh,Patra:2024eke} and bulk
acoustic wave devices~\cite{Goryachev:2014yra}, respectively.
The data points are taken from Ref.~\cite{Aggarwal:2025noe}.
Searches using ion traps can probe frequency bands that remain
largely unexplored by existing experiments.
Under the magnetic field assumed in Eq.~\eqref{eq:GWcom}, the indirect
detection via the center-of-mass mode provides better sensitivity than the
direct search using the stretch mode.
Nevertheless, searches based on the stretch mode remain valuable for
discriminating gravitational wave signals from those induced by
wave-like dark matter.


\bibliography{bibcollection}
\bibliographystyle{modifiedJHEP}


\end{document}